\documentclass[conference]{IEEEtran}
\usepackage{amsmath}
\usepackage{multirow}
\usepackage{color}
\usepackage{algorithm}
\usepackage{algpseudocode}
\usepackage{graphicx}
\usepackage{pifont}
\usepackage{booktabs}
\usepackage{siunitx}
\newcommand{\cmark}{\ding{51}}
\usepackage{fancyhdr}

\begin{document}

%
\title{Community Detection in Political Twitter Networks using Nonnegative Matrix Factorization Methods}

\author{\IEEEauthorblockN{\large Mert Ozer}
\IEEEauthorblockA{School of Computing, Informatics, \\
Decision Systems Engineering\\
Arizona State University\\
Tempe, USA 85287\\
mozer@asu.edu}
\and
\IEEEauthorblockN{\large Nyunsu Kim}
\IEEEauthorblockA{School of Computing, Informatics, \\
Decision Systems Engineering\\
Arizona State University\\
Tempe, USA 85287\\
nkim30@asu.edu}
\and
\IEEEauthorblockN{\large Hasan Davulcu}
\IEEEauthorblockA{School of Computing, Informatics, \\
Decision Systems Engineering\\
Arizona State University\\
Tempe, USA 85287\\
hdavulcu@asu.edu}}


%



\maketitle
\begin{abstract}
Community detection is a fundamental task in social network analysis. In this paper, first we develop an endorsement filtered user connectivity network by utilizing Heider's structural balance theory and certain Twitter triad patterns. Next, we develop three Nonnegative Matrix Factorization frameworks to investigate the contributions of different types of user connectivity and content information in community detection. We show that user content and endorsement filtered connectivity information are complementary to each other in clustering politically motivated users into pure political communities. Word usage is the strongest indicator of users' political orientation among all content categories. Incorporating user-word matrix and word similarity regularizer provides the missing link in connectivity-only methods which suffer from detection of artificially large number of clusters for Twitter networks.
\end{abstract}


%
\IEEEpeerreviewmaketitle

\section{Introduction}
Twitter has become one of the main stages of political activity both among politicians and partisan crowds. We have seen huge political mobilizations over Twitter in recent uprisings such as the Arab Spring and the Gezi protests \cite{arab}. Since then, politicians have been engaging in using Twitter to attract supporters and people have been using it to express their political views and opinions on various leaders and issues.

Community detection is a fundamental task in social network analysis \cite{girvan}. A community \cite{Pei:tri} can be defined as a group of users that (1) interact with each other more frequently than with those outside the group and (2) are more similar to each other than to those outside the group. Utilizing community detection algorithms to detect online political camps has attracted many researchers \cite{tang:comm, sachan, ruan}. In this work, we propose three nonnegative matrix factorization frameworks to exploit both user connectivity and content information in Twitter to find ideologically pure communities in terms of their members' political orientations.

Twitter presents three types of connectivity information between users: follow, retweet and user mention. In this paper, we do not use follow information since follow relationships correspond to longer-term structural bonds \cite{myers} and it remains challenging to determine if a follow relationship between a pair of users indicate political support or opposition. Furthermore, it has been observed that neither user retweets nor user mentions always indicate endorsement in Twitter \cite{tufekci}. However in the political sub-domain of Twitter, it has been shown that retweets tend to happen between like-minded users rather than between members of opposing camps \cite{conover}.

Using both connectivity and content information for community detection in social networks has been a popular approach among many researchers' prior works \cite{tang:comm}, \cite{sachan}, \cite{ruan}, \cite{Pei:tri}. In \cite{tang:comm}, Tang et al. propose a general framework for integrating multiple heterogenous data sources for community detection. Tang's work does not pay attention to identifying the endorsement subgraph of the connectivity graph. In \cite{sachan} Sachan et al. propose an LDA-like social interaction model by representing user connectivity as a document alongside message content. This approach also does not discriminate between positive or negative user engagement. In \cite{ruan}, Ruan et al. propose to use a filtered graph to eliminate ambiguous interactions by checking content similarity in the user's neighborhood. In this formulation, only local content patterns are taken into consideration whereas in our formulations we incorporate the global content patterns into our optimization framework.

The contributions of this paper can be summarized as follows:
\begin{itemize}
\item We start with \emph{retweets without edits} as indicators of positive endorsements between users and utilize Heider's P-O-X triad balance theory \cite{heider:tsb} to incorporate selected "structurally balanced'' \emph{edited retweets} and \emph{user mentions} into a weighted undirected connectivity graph as additional indicators of positive endorsements.
\item We develop algorithms which incorporate users' content information in our community detection frameworks to overcome the sparse nature of Twitter connectivity networks. We break down Twitter message content into three categories; words, hashtags and urls, and design experiments to measure the performance contributions of each category. Proposed Nonnegative Matrix Factorization (NMF) algorithms use user-word, user-hashtag and user-domain frequency matrices to be factorized into lower rank user vector representations while regularizing over user connectivity and content similarity to map users into their respective communities.
\end{itemize}

Pei et al. in \cite{Pei:tri} also model the problem as nonnegative matrix tri-factorization problem which factorizes user-word, tweet-word and user-user matrices into lower rank representations of users and tweets while regularizing it with user interaction and message similarity matrices. They build user-user connectivity matrix by utilizing the structural follow relationships which do not capture dynamic political context-sensitive engagement. They treat all user mentions and retweets identically and without any discrimination for endorsement. Their framework also lacks word similarity regularization.

We develop and experiment with three nonnegative matrix factorization frameworks: MultiNMF, TriNMF, DualNMF, which incorporate connectivity alongside different types of content information as regularizers. After experimenting with different dimensions of user content and different types of induced connectivity networks we discovered that incorporating more information does not necessarily yield higher clustering performance. Highest quality clustering is achieved through endorsement filtered connectivity based on methods we develop in Section \ref{sec:tsb} alongside user-word matrix based content regularization. Our DualNMF framework gives purity scores around 88\%, adjusted rand index around 75\% and NMI around 67\%. It improves all of the other baseline methods significantly as presented in Section \ref{sec:exps} and it also improves over the NMTF framework developed recently by Pei et al. \cite{Pei:tri} by 8\% in purity, 47\% in ARI and  by up to 60\% in NMI metrics. Proposed endorsement filtered sub-graph of user mentions and retweets also improves all baseline methods in almost all of the experimental setups by up to 109\% in NMI, 71\% in ARI and 17\% in purity.

The rest of the paper is organized as follows. Section \ref{sec:RelatedWork} briefly surveys related work. In Section \ref{sec:tsb}, we present Heider's theory of P-O-X structural balance of triads and its application to retweet and mention graphs to identify endorsement filtered user connectivity networks. In Section \ref{sec:ProposedMethods}, we introduce our three nonnegative matrix factorization frameworks for community detection. In Section \ref{sec:exps}, we present our experimental design, evaluation metrics and results. Section \ref{sec:Conclusion} concludes the paper and discusses future work.
\section{Related Work}
\label{sec:RelatedWork}
\subsection{Community Detection}
Since the introduction of the modularity metric by Newman in \cite{Newman:Mod}, plenty of modularity based community detection methods have been proposed in the literature \cite{Fortunato:comm,Blondel:Louvain,Clauset:CNM,Waltman:SLM}. We employ Blondel et al. \cite{Blondel:Louvain} and Clauset et al. \cite{Clauset:CNM} works as baseline algorithms to compare with ours due to their wide popularity among practitioners. A general drawback of these algorithms, when they are applied to Twitter networks, is that due to the sparse nature of the connectivity they end up with an artificially large number of communities.
\subsection{Nonnegative Matrix Factorization}
Nonnegative Matrix Factorization(NMF) algorithms by Lee et al. \cite{Lee:first} and Lin et al. \cite{Lin:firstCorr} have been extensively used and extended for different variations of community detection problems. Cai et al. \cite{cai:gnmf} introduced GNMF algorithm to incorporate Laplacian graph regularization to the standard NMF algorithm which assumes data points are sampled from a Euclidian space which is not the case usually for real-world applications. Gu et al. \cite{gu:localreg} further incorporate local learning regularization to NMF which assumes that geometrically neighboring data points are similar to each other, and should be in the same cluster. For co-clustering purposes Ding et al. \cite{Ding:tri} propose nonnegative matrix tri-factorization with orthogonality constraints. Shang et al. introduce graph dual regularized NMF algorithm in \cite{Shang:graphdual} by claiming that not only observed data but also features lie on a manifold. Yao et al. \cite{Tong:dual} apply the same logic for collaborative filtering domain and propose a dual regularized one-class collaborative filtering method.
\section{The Structural Balance of Retweet and Mention Graph}
\label{sec:tsb}
Since P-O-X triad balance theory proposed by Heider in \cite{heider:tsb}, structural balance of signed networks has been studied extensively. Heider proposed that in a signed triad, only two combinations of eight possible sign configurations are possible for a triad to be structurally balanced. Those are the following cases;
\begin{enumerate}
\item three positive edges,
\item one positive and a pair of negative edges.
\end{enumerate}
In other words, there cannot be any structurally balanced triad having only one negative edge. We adopt this social theory for the Twitter user connectivity networks, by assuming that "retweets without edits'' imply political endorsement or an unambiguous positive edge \cite{wong:rt}. However, when a retweet is edited, it has already been shown that \cite{boyd:rt}, it does not necessarily mean endorsement anymore. Moreover, user mentions do not imply endorsements either. For these reasons, we only consider retweets without edits as positive edges. For the rest of the user actions, corresponding to retweets with edits and users mentions, it is hard to detect positivity or negativity of the edges.

\begin{center}
\begin{figure}[h]
\centering
\includegraphics[scale = 0.65]{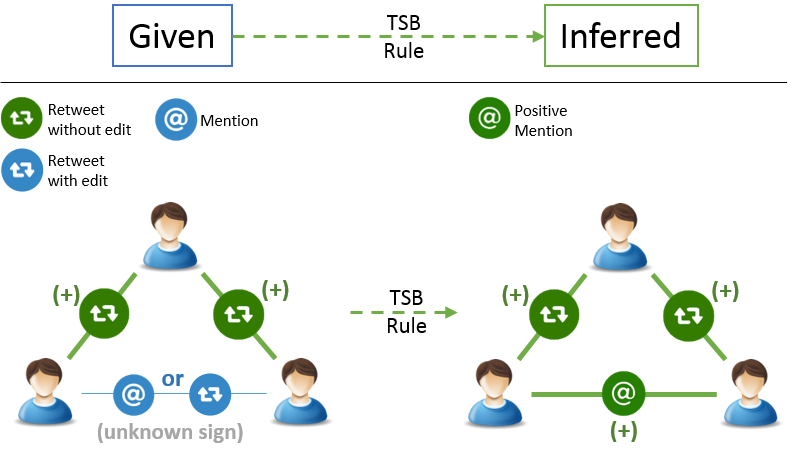}
\caption{An example application of TSB Rule}
\label{fig:triangle}
\end{figure}
\end{center}

In certain triad configurations, retweets with edits and user mentions can be identified as positive edges with the help of Heider's triad structural balance (TSB) rules. Since we do not have unambiguous negative edges, the second case is not applicable. However, since we have some positive edges to begin with, we can employ Heider's first case (i.e. three positive edges), to infer that in the presence of a triad with a pair of positive edges, the third edge can also be labeled as positive. An example configuration with a pair of positive edges is shown in Figure \ref{fig:triangle}. In this case, TSB rule is applicable and would allow us to infer that any user mention or retweet with an edit edge connecting the lower pair of users in the triad is indeed a positive edge. By employing this inference mechanism we identify the endorsement filtered user connectivity network.
\section{Proposed Methods}
\label{sec:ProposedMethods}
We propose three methods for clustering politically motivated users in Twitter namely; MultiNMF, TriNMF and DualNMF. For MultiNMF method we use document term representation of user-word, user-hashtag and user-domain matrices to be factorized and regularize the factorization problem with the user connectivity graph, cosine similarity matrices of words, domains and hashtag co-occurence matrix. For TriNMF method we use only user-word and one of user-hashtag or user-domain matrices and regularize over user connectivity and cosine domain similarity or hashtag co-occurence matrix. For DualNMF method we factorize user-word matrix into two nonnegative lower rank matrices while regularizing it with user connectivity and cosine word similarity. Before going into the details of the three algorithms we present notation in Table \ref{table:notations}.
\begin{table}[h]
\begin{center}
  \caption{Notation}
  \begin{tabular}{ | l | l | l |}
    \hline
    $\mathbf{X}_{uw}$ & user x word & frequencies of words used by users\\ \hline
    $\mathbf{X}_{uh}$ & user x hashtag & frequencies of hashtags used by users\\ \hline
    $\mathbf{X}_{ud}$ & user x domain & \begin{tabular}{@{}l@{}}frequencies of distinct domains used\\ by users\end{tabular}\\ \hline
    $\mathbf{R}$ & user x user & \begin{tabular}{@{}l@{}}adjacency matrix of \\retweet without edit graph\end{tabular}\\ \hline
    $\mathbf{M}$ & user x user & \begin{tabular}{@{}l@{}}adjacency matrix of \\mention and retweet with edit graph\end{tabular}\\ \hline
    $\Delta \mathbf{M}$ & user x user & \begin{tabular}{@{}l@{}} adjacency matrix of mentions\\ and retweet with edits completing \\retweet without edit triads\end{tabular}\\ \hline
    $\Delta \mathbf{M}_w$ & user x user & \begin{tabular}{@{}l@{}} adjacency matrix of mentions \\ and retweet with edits completing \\ retweet without edit triads \\weighted by retweet without edit edges\end{tabular} \\ \hline
    $\mathbf{C}$ & user x user & \begin{tabular}{@{}l@{}}any combination of user connectivity \\graphs\end{tabular}\\ \hline
    $\mathbf{H_{sim}}$ & hashtag x hashtag & hashtag co-occurence matrix \\ \hline
    $\mathbf{D_{sim}}$ & domain x domain & domain similarity matrix \\ \hline
    $\mathbf{W_{sim}}$ & word x word & word similarity matrix\\ \hline
    $\alpha$ & number & user connectivity regularizer parameter\\ \hline
    $\gamma$ & number & hashtag similarity regularizer parameter\\ \hline
    $\theta$ & number & domain similarity regularizer parameter\\ \hline
    $\beta$ & number & word similarity regularizer parameter\\ \hline
    $\mathbf{U}$ & user x cluster & cluster assignment matrix of users\\ \hline
    $\mathbf{H}$ & hashtag x cluster & cluster assignment matrix of hashtags\\ \hline
    $\mathbf{D}$ & domain x cluster & cluster assignment matrix of domains\\ \hline
    $\mathbf{W}$ & word x cluster & cluster assignment matrix of words\\ \hline
  \end{tabular}
  \label{table:notations}
\end{center}
\end{table}
In this work, instead of using only full user retweet and mention network we offer  three types of user connectivity regularizers as follows;
\begin{itemize}
\item $\mathbf{R} + \mathbf{M}$: It is the adjacency matrix of the full retweet and mention graph. If there exists both retweet and mention edges between two users, weights are summed up.
\item $\mathbf{R} + \Delta \mathbf{M}$: It is the adjacency matrix of the union of retweet and mention graphs in which mention edges and retweet with edits either complete a missing link in a triad of retweet without edit or already correspond to a retweet without edit edge. $\Delta \mathbf{M}$ can be formally defined as;
\begin{center}
$\Delta \mathbf{M} = \{(i,j,\mathbf{M}_{ij})\, \mid \mathbf{R}_{ij}>0 \lor \sum_{k=1}^{N}\mathbf{R}_{ik}\mathbf{R}_{kj} > 0 \} $
\end{center}
\item $\mathbf{R} + \Delta \mathbf{M}_w$: It is the adjacency matrix of the union of retweet and mention graphs in which mention edges and retweet with edits either complete a missing link in a triad of retweet without edit or already correspond to a retweet without edit edge. The ones that complete a missing link in a triad of retweet without edit are weighted by the multiplication of the weights of two retweet without edit edges in the triad. $\Delta \mathbf{M}_w$ can be defined formally as;
\begin{center}
\small$\Delta \mathbf{M}_w =\{(i,j,\mathbf{M}_{ij}(\mathbf{R}_{ij} + \sum_{k=1}^{N}\mathbf{R}_{ik}\mathbf{R}_{kj}))\} $
\end{center}
\end{itemize}

For word similarity and domain similarity regularizers we make use of cosine similarity. It can be formally defined as;\\
\begin{center}
$cos(\theta) = \dfrac{v_i \cdot v_j}{\parallel v_i \parallel * \parallel v_j \parallel}$\\
\end{center}
where $v_i$ is the user usage frequency vector of $i$th word or domain. For hashtag similarity we build similarity matrix by making use of co-occurences of hashtags in tweets. If two hashtags occur in the same tweet, we assume that those two hashtags are similar.

\subsection{MultiNMF with multi regularizers}
\label{sec:multiNMF}
To incorporate usage of both hashtags and domains of shared url links by users, we propose an NMF framework which has the following objective function;
\begin{equation}
\begin{split}
\mathbf{J_{U,H,D,W}} & =  \parallel \mathbf{X}_{uw} - \mathbf{UW}^T \parallel_F^2 + \parallel \mathbf{X}_{uh} - \mathbf{UH}^T \parallel_F^2 \\ &+ \parallel \mathbf{X}_{ud} - \mathbf{UD}^T \parallel_F^2  + \alpha Tr(\mathbf{U}^TL_\mathbf{C}\mathbf{U}) \\ &+ \gamma Tr(\mathbf{H}^TL_{\mathbf{H}_{sim}}\mathbf{H}) + \theta Tr(\mathbf{D}^TL_{\mathbf{D}_{sim}}\mathbf{D}) \\&+ \beta Tr(\mathbf{W}^TL_{\mathbf{W}_{sim}}\mathbf{W}) \\&s.t. \quad \mathbf{U} \geq 0, \mathbf{H} \geq 0, \mathbf{D} \geq 0, \mathbf{W} \geq 0
\end{split}
\label{eq:obj1}
\end{equation}
where $L_\mathbf{C}$ is the Laplacian matrix of adjacency matrix of user connectivity graph defined as $D_\mathbf{C} - \mathbf{C}$ and $D_\mathbf{C}$ is the matrix which contains the degree of each user node in its diagonals.  $L_{\mathbf{H}_{sim}}$, $L_{\mathbf{D}_{sim}}$ and $L_{\mathbf{W}_{sim}}$ follow the same definition for hashtags and words. Due to the very fuzzy multi-class nature of words, hashtags and domain names, we do not include orthogonality constraints for matrices $\mathbf{U},\mathbf{H},\mathbf{D},\mathbf{W}$, which usually result in more precise clusters for co-clustering tasks. It is easy to see that the proposed objective function is not convex for $\mathbf{U},\mathbf{H},\mathbf{D}$ and $\mathbf{W}$, hence we develop an iterative algorithm which tries to find a local minima by updating each matrix iteratively as follows; 
\begin{equation}
\mathbf{U} \gets \mathbf{U} \odot \sqrt{\dfrac{\mathbf{X}_{uw}\mathbf{W} + \mathbf{X}_{uh}\mathbf{H} + \mathbf{X}_{ud}\mathbf{D} + \alpha L_\mathbf{C}^-\mathbf{U}}{\mathbf{UW}^T\mathbf{W} + \mathbf{UH}^T\mathbf{H} + \mathbf{UD}^T\mathbf{D} + \alpha L_\mathbf{C}^+\mathbf{U}}}
\label{eq:U1}
\end{equation}
\begin{equation}
\mathbf{H} \gets \mathbf{H} \odot \sqrt{\dfrac{\mathbf{X}_{uh}^T\mathbf{H} + \gamma L_{\mathbf{H}_{sim}}^-\mathbf{H}}{\mathbf{HU}^T\mathbf{U} + \gamma L_{\mathbf{H}_{sim}}^+\mathbf{H}}}
\label{eq:H1}
\end{equation}
\begin{equation}
\mathbf{D} \gets \mathbf{D} \odot \sqrt{\dfrac{\mathbf{X}_{ud}^T\mathbf{D} + \theta L_{\mathbf{D}_{sim}}^-\mathbf{D}}{\mathbf{DU}^T\mathbf{U} + \theta L_{\mathbf{D}_{sim}}^+\mathbf{D}}}
\label{eq:D1}
\end{equation}
\begin{equation}
\mathbf{W} \gets\mathbf{W} \odot \sqrt{\dfrac{\mathbf{X}_{uw}^T\mathbf{U} + \beta L_{\mathbf{W}_{sim}}^-\mathbf{W}}{\mathbf{WU}^T\mathbf{U} + \beta L_{\mathbf{W}_{sim}}^+\mathbf{W}}}
\label{eq:W1}
\end{equation}
where $L_{ij}^+ = (|L_{ij}| + L_{ij})/2$ and $L_{ij}^- = (|L_{ij}| - L_{ij})/2$.  $\odot$ represents element-wise multiplication and $\dfrac{[\cdot]}{[\cdot]}$ represents element-wise division. Derivation of update rules can be seen in Appendix \ref{derivation}. Complexity of the method can be inferred as $\mathcal{O}(i(uwk + uhk + udk + u^2k + h^2k + d^2k + w^2k))$ when complexity of multiplying any $X$ matrix with any of $\mathbf{U},\mathbf{H},\mathbf{D},\mathbf{W}$ is considered to be $\mathcal{O}(uwk)$, $\mathcal{O}(uhk)$, $\mathcal{O}(udk)$ and multiplying any of Laplacian matrices $L$ with any of $\mathbf{U},\mathbf{H},\mathbf{D},\mathbf{W}$ is taken as $\mathcal{O}(u^2k)$, $\mathcal{O}(h^2k)$, $\mathcal{O}(d^2k)$ or $\mathcal{O}(w^2k)$ where $i$ is the number of iterations, $u$ is number of users, $h$ is the number of hashtags, $d$ is the number of domains, $w$ is the number of words and  $k$ is the number of clusters. The general algorithmic framework is given at the end of methodology in Algorithm \ref{NMF Algorithms}.
\subsection{TriNMF with three regularizers}
To incorporate usage of hashtags or domains of shared url links solely, we propose a new NMF framework which has the following objective function.
\begin{equation}
\begin{split}
\mathbf{J_{U,H,W}} &=  \parallel \mathbf{X}_{uw} - \mathbf{UW}^T \parallel_F^2 + \parallel \mathbf{X}_{uh} - \mathbf{UH}^T \parallel_F^2 \\ &+ \alpha Tr(\mathbf{U}^TL_\mathbf{C}\mathbf{U}) + \gamma Tr(\mathbf{H}^TL_{\mathbf{H}_{sim}}\mathbf{H}) \\ &+ \beta Tr(\mathbf{W}^TL_{\mathbf{W}_{sim}}\mathbf{W}) \\&s.t. \quad \mathbf{U} \geq 0, \mathbf{H} \geq 0, \mathbf{W} \geq 0
\end{split}
\end{equation}
where $L_\mathbf{C}$ is the Laplacian matrix of user connectivity defined as $D_\mathbf{C} - \mathbf{C}$ and $D_\mathbf{C}$ is a diagonal matrix which contains the degree of each user in its diagonals.  $L_{\mathbf{H}_{sim}}$ and $L_{\mathbf{W}_{sim}}$ follows the same definition for hashtags and words. After applying the same procedure followed in Section \ref{sec:multiNMF}, we get updating rules as follows.
\begin{equation}
\mathbf{U} \gets \mathbf{U} \odot \sqrt{\dfrac{\mathbf{X}_{uw}\mathbf{W} + \mathbf{X}_{uh}\mathbf{H} + \alpha L_\mathbf{C}^-\mathbf{U}}{\mathbf{UW}^T\mathbf{W} + \mathbf{UH}^T\mathbf{H} + \alpha L_\mathbf{C}^+\mathbf{U}}}
\label{eq:U2}
\end{equation}
\begin{equation}
\mathbf{H} \gets \mathbf{H} \odot \sqrt{\dfrac{\mathbf{X}_{uh}^T\mathbf{U} + \gamma L_{\mathbf{H}_{sim}}^-\mathbf{H}}{\mathbf{HU}^T\mathbf{U} + \gamma L_{\mathbf{H}_{sim}}^+\mathbf{H}}}
\label{eq:H2}
\end{equation}
\begin{equation}
\mathbf{W} \gets \mathbf{W} \odot \sqrt{\dfrac{\mathbf{X}_{uw}^T\mathbf{U} + \beta L_{\mathbf{W}_{sim}}^-\mathbf{W}}{\mathbf{WU}^T\mathbf{U} + \beta L_{\mathbf{W}_{sim}}^+\mathbf{W}}}
\label{eq:W2}
\end{equation}
 Note that this update rules can be obtained by setting $\mathbf{D}$, $\mathbf{D}_{sim}$ and $\theta$ equal to 0 in Equations \ref{eq:U1}, \ref{eq:H1}, \ref{eq:W1}. Complexity of the method can be calculated by omitting the costs of operations done over matrices $\mathbf{X}_{ud}$, $\mathbf{D}$ and $L_{\mathbf{D}_{sim}}$. The complexity of the method is $\mathcal{O}(i(uwk + uhk + u^2k + h^2k + w^2k))$.
\subsection{DualNMF}
To use only user word matrix as user content and regularize factorization with user connectivity and keyword similarity, inspired by \cite{Tong:dual}, we present DualNMF objective function as;
\begin{equation}
\begin{split}
\mathbf{J_{U,W}} &=  \parallel \mathbf{X}_{uw} - \mathbf{UW}^T \parallel_F^2 + \alpha Tr(\mathbf{U}^TL_\mathbf{C}\mathbf{U}) \\ &+ \beta Tr(\mathbf{W}^TL_{\mathbf{W}_{sim}}\mathbf{W}) \\&s.t. \quad \mathbf{U} \geq 0, \mathbf{W} \geq 0
\end{split}
\end{equation}
After following the same procedure introduced in Section \ref{sec:multiNMF}, we can get the update rules for $\mathbf{U}$ and $\mathbf{W}$ as;
\begin{equation}
\mathbf{U} \gets \mathbf{U} \odot \sqrt{\dfrac{\mathbf{X}_{uw}\mathbf{W} + \alpha L_\mathbf{C}^-\mathbf{U}}{\mathbf{UW}^T\mathbf{W} + \alpha L_\mathbf{C}^+\mathbf{U}}}
\label{eq:U3}
\end{equation}
\begin{equation}
\mathbf{W} \gets \mathbf{W} \odot \sqrt{\dfrac{\mathbf{X}_{uw}^T\mathbf{U} + \beta L_{\mathbf{W}_{sim}}^-\mathbf{W}}{\mathbf{WU}^T\mathbf{U} + \beta L_{\mathbf{W}_{sim}}^+\mathbf{W}}}
\label{eq:W3}
\end{equation}
Complexity of the method can be inferred as $\mathcal{O}(i(uwk + u^2k + w^2k))$ after omitting the extra operations done over matrices $\mathbf{X}_{uh}$, $\mathbf{H}$ and $D_{\mathbf{H}_{sim}}$ in the previous method.        
\begin{algorithm}
\caption{NMF Algorithms}
\label{NMF Algorithms}
\begin{algorithmic}[1]
\Statex\textbf{Input}:$\{\mathbf{X}_{uw},\mathbf{X}_{uh},\mathbf{X}_{ud},\mathbf{C},\mathbf{H}_{sim},\mathbf{D}_{sim},\mathbf{W}_{sim},\alpha,\beta,\theta,\gamma\}$
\Statex\textbf{Output}: $\mathbf{U}$
\State Initialize $\mathbf{U},\mathbf{H},\mathbf{D},\mathbf{W} > 0$

\While{$\Delta residual > threshold$}
\State Update $\mathbf{U}$ by using one of Equations \ref{eq:U1}, \ref{eq:U2}, \ref{eq:U3}
\State Update $\mathbf{H}$ by using one of Equations \ref{eq:H1}, \ref{eq:H2}
\State Update $\mathbf{D}$ by using Equation \ref{eq:D1}
\State Update $\mathbf{W}$ by using one of Equations \ref{eq:W1}, \ref{eq:W2}, \ref{eq:W3}
\EndWhile
\State Assign user i to community j where $j = argmax_j{\mathbf{U}_{ij}}$.
\end{algorithmic}
\end{algorithm}
The general algorithm can be summarized as the application of the related update rules to the matrices $\mathbf{U},\mathbf{H},\mathbf{D},\mathbf{W}$. For MultiNMF with multi regularizers method, equations \ref{eq:U1}, \ref{eq:H1}, \ref{eq:D1}, \ref{eq:W1} are applied. For TriNMF with three regularizers method, equations \ref{eq:U2}, \ref{eq:H2}, \ref{eq:W2} are applied and $\mathbf{D}$ matrix is not included in calculations. For DualNMF method, equations \ref{eq:U3} and \ref{eq:W3} are applied and $\mathbf{H}$ and $\mathbf{D}$ matrices are not included in calculations.
\section{Experiments and Results}
\label{sec:exps}
\subsection{Data Description}
We make use of a pair of publicly available\footnote{Users' Twitter id lists can be obtained from http://mlg.ucd.ie/aggregation/index.html} political Twitter datasets to evaluate our methods. These datasets are user lists of 419 British political figures from four major political parties in the UK, namely; Conservative and Unionist Party, Labour Party, Scottish National Party, Liberal Democrats and others, and 349 major Irish political figures from seven political parties; Fianna Fáil, Fine Gael, Green Party, Sinn Féin, United Left Alliance, Independents. Several statistics for the datasets are shown in Table \ref{table:atts}.

\begin{table}[h]
\begin{center}
  \caption{Data Attributes}
	\begin{tabular}{l s s} \toprule
     & UK & Ireland \\ \midrule
    \# of Tweets & 19,947 & 14,656  \\
    \# of Retweets & 1,566 & 7,088 \\ 
    \# of Mentions & 4,956 & 22,072 \\ 
    \# of Words & 10,766 & 7,973 \\ 
    \# of Hashtags & 945 & 986 \\ 
    \# of URL Domains & 946 & 634  \\
    \# of Users & 233 & 258 \\ 
    \# of Baseline Communities & 5 & 7 \\ \bottomrule
	\end{tabular}
\label{table:atts}
\end{center}
\end{table}

For the UK and Ireland data, we crawl all of the tweets sent from the accounts of given user id lists. In order not to be heavily influenced by the extremely polarized election season, we only used tweets dated after May, 7 2015, which was the election day in the UK. To balance the share of number of tweets from each user we limit the number of tweets to 200 per user.

For each dataset, same preprocessing method is followed. First, words occurring less than 20 times and stop words are eliminated. After eliminating word features, users and tweets that lack content are also eliminated. Hashtags and domains that appear only once are not taken into consideration either. Statistics shown in Table \ref{table:atts} show the numbers after preprocessing.
\subsection{Evaluation Metrics}
To evaluate the methods, we make use of three well known clustering quality metrics, namely; purity, adjusted rand index and normalized mutual information.

\textbf{Purity} can be formally defined as;
$$ Purity = \frac{1}{n} \sum_{i=1}^{k} max_j |C_i \cap l_j| $$
where $k$ is the number of communities found, $n$ is the number of instances, $l_j$ is the set of instances which belong to the class j, and $C_i$ is the set of instances that are members of community i.

\textbf{Adjusted Rand Index} \cite{hubert:ari} can be formally defined as;
$$ ARI = \dfrac{RI - E[RI]}{max(RI) - E[RI]} $$
where
$$ RI = \dfrac{s + s'}{\binom n2} $$
$s$ is the number of pairs which belong to both same ground-truth class and identified community. $s'$ is the number of pairs which belong to both different ground-truth classes and identified communities. It evaluates the similarity of ground-truth class labels and clustering result.

\textbf{Normalized Mutual Information} can be formally defined as;
$$ NMI = \dfrac{\sum^{|l|}_{j=1}\sum^{|C|}_{i=1}P(j,i)log\Big(\dfrac{P(j,i)}{P(i)P(j)}\Big)}{\sqrt{H(l)H(C)}} $$
where, $H(l)$ and $H(C)$ are the entropy of class and community assignments of l and C. $P(j,i)$ is the probability that randomly picked user has class label j and belongs to the community i while $P(j)$ gives the probability of randomly picked user to be in class j and similarly $P(i)$ to be in community i.
\subsection{Baseline Algorithms}
As a baseline to evaluate the performance of using both connectivity and content information, we design experiments with connectivity-only and content-only clustering methods.

For connectivity-only method, we use Louvain \cite{Blondel:Louvain} and CNM \cite{Clauset:CNM} algorithms utilizing modularity optimization over user adjacency matrix. Modularity is defined as:
\begin{equation}
	Q = \frac{1}{2m}\sum_{ij} (A_{ij} - \dfrac{k_ik_j}{2m}) \delta (c_{i},c_{j})
	\label{eq:modularity1}
\end{equation}
where $\delta (c_{i},c_{j})$ is the Kronecker delta symbol, $c_{i}$ is the label of the community to which node i is assigned, and $k_i$ is the degree of node i.

For content-only approach, we experiment with k-means\cite{kmeans} and conventional non-negative matrix factorization algorithm \cite{Lee:first}.

For approaches employing both connectivity and content information of users, we test GNMF \cite{cai:gnmf} and NMTF \cite{Pei:tri} algorithms besides proposed methods. GNMF algorithm is introduced by Cai et al. to incorporate intrinsic geometric similarity of users. We feed previously defined three types of user connectivity graphs' adjacency matrices as graph regularization terms to the GNMF algorithm.

Pei et al. work in \cite{Pei:tri} applies nonnegative matrix tri-factorization with regularization to Twitter data. It makes use of user similarity, [tweet x word] and [user x word] matrices and regularize the objective function with tweet similarity and user connectivity matrices. Complexity of the algorithm is $O(rk(mn + mw + nw + m^3 + n^2))$ where r is the iteration times. m, n, k, and w denote the number of users, messages, features and communities.

\subsection{Experimental Design}
First set of experiments test the performance of using connectivity-only information for community detection, labeled as the Experiment Set 1. We test Louvain and CNM algorithms on three different types of connectivity graphs. 
Second set of experiments test the performance of content-only methods, labeled as Experiment Set 2. We test k-means and NMF methods. Third set of experiments test the performance of methods utilizing both connectivity and content information, labeled as Experiment Set 3. We test GNMF and NMTF frameworks proposed by \cite{Pei:tri} as baseline algorithms, alongside our proposed MultiNMF, TriNMF and DualNMF methods. In user content dimension, we use DualNMF method to test the experiment design that only uses user-word content. We use TriNMF method to test the experiment design that uses user-hashtag or user-domain information in combination with the user-word information. We use MultiNMF method to test the experiment design that uses all of user-word, user-hashtag and user-domain contents. We label these experiments as Experiment Set 3.1, 3.2 and 3.3 respectively.

\subsection{Experimental Results}

First, we present statistics of retweets without edits and user mentions on the full and endorsement filtered user connectivity graphs. Table \ref{table:innerinter} shows that retweeting without edits indeed occurs mostly inside like-minded political camps, rather than cross-camps. Roughly 97\% of retweets in the UK data, and 88\% of retweets in the Ireland data occur inside like-minded groups, while these percentages are much lower for users mentions. Our endorsement filtered connectivity network boosts the percentage of inner group user mentions from 83\% to 97\% in the UK data and from 59\% to 87\% in the Ireland data evidencing that TSB rule in fact identifies positive user mentions and retweets with edits with high accuracy.

\begin{table}[h]
\begin{center}
  \caption{Data Attributes}
	\begin{tabular}{l s s} \toprule
     & UK & Ireland \\ \midrule
    Inner Group Retweet Links & 962 & 1,652  \\
    Inter Group Retweet Links & 28 & 216 \\ \midrule
    Inner Group Retweet + Mention Links & 1,986 & 3,056 \\ 
    Inter Group Retweet + Mention Links & 398 & 2,092 \\ \midrule
    Inner Group Retweet + $\Delta$Mention Links & 1,456 & 2,820 \\ 
    Inter Group Retweet + $\Delta$Mention Links & 40 & 432  \\ \bottomrule
	\end{tabular}
\label{table:innerinter}
\end{center}
\end{table}

%

  We run each experiment 20 times for every method and pick the maximum score achieved for reporting. Each regularizer parameter ($\alpha$, $\gamma$, $\theta$, $\beta$) are experimented with values 1, 10, 100 and 1000. Best accuracies are usually reached with experiments in which $\alpha$ and $\beta$ equal to 10 or 100 while $\gamma$ and $\theta$ equal to 1. This shows the contribution of user connectivity and word similarity regularizers, and considerably lower contributions of hashtag and domain name regularizers towards overall performance of the algorithms.
\begin{table}[h]
\begin{center}
  \caption{UK Experiment Set 1 Results}
  \begin{tabular}{  l  l  c  c  c c  c } \toprule
	Algorithm & User Graph & k & Purity & ARI & NMI \\ \midrule
	\multirow{3}{*}{Louvain} & $R + M$ & 20 & .9313 & .4661 & .5854 \\ 
	& $R +\Delta M$ & 42 & .9613 & .3691 & .5916 \\ 
	& $R +\Delta M_{w}$ & 42 & .9484 & .4291 & .5916 \\ \midrule
	\multirow{3}{*}{CNM} & $R +M$ & 17 & .8498 & .5656 & .5257 \\ 
	& $R +\Delta M$ & 41 & .9700 & .6150 & .6496 \\ 
	& $R +\Delta M_{w}$ & 41 & .9700 & .6150 & .6496 \\ \bottomrule
	\end{tabular}
  \label{table:ukConnectivity}
\end{center}
\end{table}
\begin{table}[h]
\begin{center}

  \caption{Ireland Experiment Set 1 Results}
  \begin{tabular}{  l  l  c  c  c c  c }\toprule
  Algorithm & User Graph & k & Purity & ARI & NMI \\ \midrule

	\multirow{3}{*}{Louvain} & $R +M$ & 13 & .8720 & .7277 & .6849 \\ 
	 & $R +\Delta M$ & 31 & .9186 & .7453 & .7393\\ 
	 & $R +\Delta M_{w}$ & 31 & .9224  & .7536 & .7518\\ \midrule
	\multirow{3}{*}{CNM} & $R +M$ & 10 & .7016 & .4509 & .4720 \\ 
	 & $R +\Delta M$ & 29 & .8333 & .6426 & .6381\\ 
	 & $R +\Delta M_{w}$ & 29 & .8333 & .6426 & .6381\\ \bottomrule
	 \end{tabular}
  \label{table:irelandConnectivity}
\end{center}
\end{table}

Major findings for Experiment Set 1 can be summarized as follows:
\begin{itemize}
\item Relatively larger clustering scores occur due to artificially large number of clusters that are found. Considering the number of users in both datasets, the number of clusters identified in Experiment Set 1 are not practical for use (e.g. 29 clusters in Ireland data for 7 political parties).
\item Using endorsement filtered user connectivity graph usually gives better clustering performance compared to using full user connectivity graph. There is a pattern of weighted graph approach outperforming the others.
\end{itemize}

\begin{table}[h]
\begin{center}
  \caption{UK Experiment Set 2 Results}
  \begin{tabular}{  l  l   c  c  c  c }\toprule
  Algorithm & User Content & Purity & ARI & NMI \\ \midrule
    k-Means & user x word & .6738  & .2378 & .2018 \\ 
    NMF & user x word & .6395 & .1541 & .1709 \\ \bottomrule
    \end{tabular}
  \label{table:ukContent}
\end{center}
\end{table}

\begin{table}[h]
\begin{center}
  \caption{Ireland Experiment Set 2 Results}
  \begin{tabular}{  l  l  c  c  c  c }\toprule
  Algorithm & User Content & Purity & ARI & NMI \\ \midrule
    k-Means & user x word & .4651 & .0488 & .1672 \\ 
    NMF & user x word & .4186 & .0434 & .1139 \\ \bottomrule
    \end{tabular}
  \label{table:irelandContent}
\end{center}
\end{table}

Experiment Set 2 indicates that word usage-only based clustering yields considerably lower accuracies compared to user connectivity-only based clustering.

\begin{table}[h]
\begin{center}
  \caption{UK Experiment Set 3 Results}
  \begin{tabular}{  l  l  l  c  c  c  c  c }\toprule
  Algorithm & User Graph & User Content & Purity & ARI & NMI \\ \midrule
    \multirow{3}{*}{GNMF\cite{cai:gnmf}} & $R +M$ & user x word & .7854 & .4955 & .4120\\ 
    & $R +\Delta M$ & & .8069 & .6099 & .4922\\ 
    & $R +\Delta M_{w}$ & & .8326 & .6469 & .5461\\ \midrule
   \multirow{3}{*}{NMTF\cite{Pei:tri}} & $R +M$ & user x word, & .8197 & .6448 & .2593\\ 
    & $R +\Delta M$ & tweet x word & .8112 & .5657 & .2471\\ 
    & $R +\Delta M_{w}$ & & .8412 & .5331 & .3751\\ \midrule
    \multirow{3}{*}{TriNMF} & $R +M$ & user x word, & .7597 & .3707 & .3158\\ 
    & $R +\Delta M$ & user x domain & .7940 & .5566 & .4595\\ 
    & $R +\Delta M_{w}$ & & .8283 & .6375 & .5006\\ \midrule
    \multirow{3}{*}{TriNMF}  & $R +M$ & user x word, & .7897 & .5232 & .4320 \\ 
    & $R +\Delta M$ & user x hashtag &  .8112 & .4640 & .3780\\ 
    & $R +\Delta M_{w}$ & & .7768 & .5001 & .3837\\ \midrule
    \multirow{3}{*}{MultiNMF}  & $R +M$ & user x word & .7554 & .4025  & .3343 \\ 
    & $R +\Delta M$ & user x domain, & .8112 & .5726 &.4404 \\ 
    & $R +\Delta M_{w}$ & user x hashtag & .8112 & .6108 & .4978\\ \midrule
    \multirow{3}{*}{DualNMF} & $R +M$ & user x word & .8326 & .5674 & .5146\\ 
    & $R +\Delta M$ & & .8927 & .7291 & .6086\\ 
    & $R +\Delta M_{w}$ & & \textbf{.8970} & \textbf{.7616} & \textbf{.6380}\\ \bottomrule
  \end{tabular}
  \label{table:ukpolitics}
\end{center}
\end{table}

\begin{table}[h]
\begin{center}

  \caption{Ireland Experiment Set 3 Results}
    \begin{tabular}{  l  l  l  c  c  c }\toprule
    Algorithm & User Graph & Content & Purity & ARI & NMI \\ \midrule
    \multirow{3}{*}{GNMF\cite{cai:gnmf}} & $R +M$ & user x word & .5543 & .2447 & .2881\\ 
     & $R +\Delta M$ & & .6279 & .4557 & .4652\\ 
     & $R +\Delta M_{w}$ & & .8178 & .6978 & .6399\\ \midrule
    \multirow{3}{*}{NMTF\cite{Pei:tri}} & $R +M$ & user x word,  & .5969 & .3119 & .2144\\ 
     & $R +\Delta M$ & tweet x word & .6860 & .3986 & .2384\\ 
     & $R +\Delta M_{w}$ & & .7597 & .5198 & .4469\\ \midrule
	\multirow{3}{*}{TriNMF} & $R +M$ & user x word, & .7209 & .5051 & .5237\\ 
    & $R +\Delta M$ & user x domain & .7946 & .6045 & .5313\\ 
    & $R +\Delta M_{w}$ & & .8101 & .6807 & .6372\\ \midrule
    \multirow{3}{*}{TriNMF}  & $R +M$ & user x word, & .6938 & .4202 & .4431 \\ 
    & $R +\Delta M$ & user x hashtag &  .7016 & .5300 & .4224\\ 
    & $R +\Delta M_{w}$ & & .8062 & .6784 & .6885\\ \midrule
    \multirow{3}{*}{MultiNMF}  & $R +M$ & user x word, & .7481 & .4777  & .4938 \\ 
    & $R +\Delta M$ & user x domain, & .6744 & .4597 & .4219 \\ 
    & $R +\Delta M_{w}$ & user x hashtag & .8178 & .6953 & .6411\\ \midrule
    \multirow{3}{*}{DualNMF} & $R +M$ & user x word & .7364 & .5561 & .5397\\ 
     & $R +\Delta M$ & & .7597 & .6292 & .6029\\ 
     & $R +\Delta M_{w}$ & & \textbf{.8721} & \textbf{.7536} & \textbf{.7096}\\ \bottomrule
  \end{tabular}
  \label{table:irelandpolitics}
\end{center}
\end{table}

\begin{table}[h]
\begin{center}
  \caption{Comparison of Methods for Experiment Set 3}
  \begin{tabular}{  l  l  | c | c |}
  \\\cline{3-4}
   &  & \multicolumn{2}{|c|}{Connectivity}  \\ \cline{3-4}
   & & $R + M$ & \cmark$\mathbf{R + \Delta M_w}$\\ \hline
    \multicolumn{1}{|c|}{\multirow{5}{*}{Content}} & \cmark \textbf{word} & DualNMF  & \cmark\cmark \textbf{DualNMF}  \\ \cline{2-4}
    \multicolumn{1}{|c|}{} & word, & \multirow{2}{*}{TriNMF}  & \multirow{2}{*}{\cmark \textbf{TriNMF}} \\
    \multicolumn{1}{|c|}{} & hashtag or domain & & \\ \cline{2-4}
   \multicolumn{1}{|c|}{} & word, & \multirow{2}{*}{MultiNMF}  & \multirow{2}{*}{\cmark \textbf{MultiNMF}}\\
    \multicolumn{1}{|c|}{} & hashtag and domain & & \\ \hline
    \end{tabular}
  \label{table:exSet3}
\end{center}
\end{table}

Major findings from Experiment Set 3 can be summarized as follows;
\begin{itemize}
	\item Regardless of the experiment set and algorithms used, endorsement filtered user connectivity graph yields higher accuracy clustering performance compared to using the full connectivity graph. Usually weighted graph approach outperforms the others.
	\item DualNMF method which factorizes user-word matrix alongside user connectivity and word similarity regularizers yields the highest accuracy clustering performance.
	\item We get much higher scores of clustering accuracy in Experiment Set 3 compared to Experiment Set 2. Regularizing content-only methods with user connectivity graphs(GNMF \cite{cai:gnmf}), dramatically increases the quality of the clustering. DualNMF which incorporates keyword similarity regularization to GNMF further boosts the quality of clustering.
 	\item Compared to DualNMF method, including tweet messages for NMTF method proposed in \cite{Pei:tri} does not help to further improve the clustering quality, while it increases complexity dramatically. DualNMF provides 9\% additional purity, 46\% additional ARI score while doubling the NMI score compared to the baseline NMTF method of Pei et al. in \cite{Pei:tri}.
  	\item Compared to DualNMF method, utilizing hashtag and/or domain usage information (i.e. TriNMF and MultiNMF) do not contribute to the overall clustering quality.
\end{itemize}

\section{Conclusion}
\label{sec:Conclusion}
In Twitter, content and endorsement filtered connectivity are complementary to each other in clustering politically motivated users into pure political communities. Word usage is the strongest indicator of user's political orientation among all content categories. Incorporating user-word matrix and word similarity regularizer provides the missing link in connectivity-only methods which suffers from detection of artificially large number of clusters in sparse Twitter networks. Our future work includes parallel distributed evolutionary community detection and identification of emerging coalitions and conflicts among communities.

\section*{Acknowledgment}
This research was supported by ONR Grants N00014-16-1-2386 and N00014-15-1-2722.




\begin{thebibliography}{1}
\bibitem{arab}
Howard, Philip N., and Aiden Duffy, Deen Freelon, Muzammil Hussain, Will Mari, and Marwa Mazaid. Opening Closed Regimes: What Was the Role of Social Media During the Arab Spring? Project on Information Technology and Political Islam Data Memo 2011.1. Seattle: University of Washington, 2011.
\bibitem{girvan}
Girvan,M., Newman M.E.J. (2002). Community structure in social and biological networks, Proceedings of the National Academy of Sciences, 99(12), pp.7821-7826.
\bibitem{Pei:tri}
Yulong Pei, Nilanjan Chakraborty, and Katia Sycara. 2015. Nonnegative matrix tri-factorization with graph regularization for community detection in social networks. In Proceedings of the 24th International Conference on Artificial Intelligence (IJCAI'15), Qiang Yang and Michael Wooldridge (Eds.). AAAI Press 2083-2089.
\bibitem{tang:comm}
J. Tang, X. Wang, and H. Liu. Integrating Social Media Data for Community Detection. In Modeling and Mining Ubiquitous Social Media, 2012.
\bibitem{sachan}
Mrinmaya Sachan, Danish Contractor, Tanveer A. Faruquie, and L. Venkata Subramaniam. 2012. Using content and interactions for discovering communities in social networks. In Proceedings of the 21st international conference on World Wide Web (WWW '12). ACM, New York, NY, USA, 331-340.
\bibitem{ruan}
Y. Ruan, D. Fuhry, and S. Parthasarathy. Efficient community detection in large networks using content and links. In WWW ’13, 2013.
\bibitem{tufekci}
Tufekci, Z. (2014). Big Questions for Social Media Big Data: Representativeness, Validity and Other Methodological Pitfalls. In: International AAAI Conference on Weblogs and Social Media.
\bibitem{conover}
M. D. Conover, B. Gonc¸alves, J. Ratkiewicz, M. Francisco, A. Flammini, and F. Menczer, “Political polarization on Twitter,” in Proceedings of the 5th InternationalConference on Weblogs and Social Media, 2011
\bibitem{bailon}
Sandra González-Bailón, Ning Wang, Alejandro Rivero, Javier Borge-Holthoefer, Yamir Moreno, Assessing the bias in samples of large online networks, Social Networks, Volume 38, July 2014, Pages 16-27, ISSN 0378-8733, http://dx.doi.org/10.1016/j.socnet.2014.01.004.
\bibitem{myers}
S. A. Myers, A. Sharma, P. Gupta, and J. Lin. Information network or social network? The structure of the Twitter follow graph. WWW Companion, 2014
\bibitem{heider:tsb}
Heider F. The psychology of interpersonal relations. New York: Wiley, 1958. 322 p.
\bibitem{wong:rt}
Felix Ming Fai Wong, Chee Wei Tan, Soumya Sen, and Mung Chiang. 2013. Quantifying political leaning from tweets and retweets. In Proceedings of the International AAAI Conference on Weblogs and Social Media (ICWSM).
\bibitem{boyd:rt}
D. Boyd, S. Golder and G. Lotan, "Tweet, Tweet, Retweet: Conversational Aspects of Retweeting on Twitter," System Sciences (HICSS), 2010 43rd Hawaii International Conference on, Honolulu, HI, 2010, pp. 1-10.
doi: 10.1109/HICSS.2010.412
\bibitem{hubert:ari}
Hubert, L.,  Arabie, P. (1985). Comparing partitions. Journal of Classification, 2, 193–218.
\bibitem{boyd:opt}
Stephen Boyd and Lieven Vandenberghe. Convex optimization. Cambridge University Press, Cambridge, 2004.
\bibitem{Newman:Mod}
M. Newman. 2006. Modularity and community structure in networks. Proceedings of the National Academy of Sciences, vol. 103, no. 23, pp. 8577–8582.
\bibitem{Fortunato:comm}
S. Fortunato. 2010. Community detection in graphs. Physics Reports, vol. 486, pp. 75–174
\bibitem{Blondel:Louvain}
Vincent D Blondel, Jean-Loup Guillaume, Renaud Lambiotte, Etienne Lefebvre. Fast unfolding of communities in large networks. Journal of Statistical Mechanics: Theory and Experiment 2008 (10), P10008 (12pp)
doi: 10.1088/1742-5468/2008/10/P10008.
\bibitem{Clauset:CNM}
A. Clauset, M. Newman, and C. Moore, “Finding community
structure in very large networks,” Physical Review E, vol. 70,
p. 066111, 2004. [Online]. Available: http://www.citebase.org/cgibin/citations?id=oai:arXiv.org:cond-mat/0408187
\bibitem{Waltman:SLM}
Waltman, L., Van Eck, N. J.. 2013. A smart local moving algorithm for large-scale modularity-based community detection.
European Physical Journal B, 86, 471.
\bibitem{kmeans}
Lloyd., S. P. (1982). "Least squares quantization in PCM" (PDF). IEEE Transactions on Information Theory 28 (2): 129–137. doi:10.1109/TIT.1982.1056489
\bibitem{Lee:first}
Daniel D. Lee, H. Sebastian Seung. 2000. Algorithms for Non-negative Matrix Factorization. In Neural Information Processing Systems (NIPS), Vol. 13 , pp. 556-562.
\bibitem{Lin:firstCorr}
C. J. Lin. On the Convergence of Multiplicative Update Algorithms for Nonnegative Matrix Factorization. In IEEE Transactions on Neural Networks, vol. 18, no. 6, pp. 1589-1596, Nov. 2007. doi: 10.1109/TNN.2007.895831
\bibitem{Tong:dual}
Yuan Yao, Hanghang Tong, Guo Yan, Feng Xu, Xiang Zhang, Boleslaw K. Szymanski, and Jian Lu. 2014. Dual-Regularized One-Class Collaborative Filtering. In Proceedings of the 23rd ACM International Conference on Conference on Information and Knowledge Management (CIKM '14). ACM, New York, NY, USA, 759-768. DOI=http://dx.doi.org/10.1145/2661829.2662042
\bibitem{Zhu:sent}
Linhong Zhu, Aram Galstyan, James Cheng, and Kristina Lerman. 2014. Tripartite graph clustering for dynamic sentiment analysis on social media. In Proceedings of the 2014 ACM SIGMOD International Conference on Management of Data (SIGMOD '14). ACM, New York, NY, USA, 1531-1542. DOI=http://dx.doi.org/10.1145/2588555.2593682
\bibitem{Gu:co-manifolds}
Quanquan Gu and Jie Zhou. 2009. Co-clustering on manifolds. In Proceedings of the 15th ACM SIGKDD international conference on Knowledge discovery and data mining (KDD '09). ACM, New York, NY, USA, 359-368. DOI=http://dx.doi.org/10.1145/1557019.1557063
\bibitem{Shang:graphdual}
Fanhua Shang, L. C. Jiao, and Fei Wang. 2012. Graph dual regularization non-negative matrix factorization for co-clustering. Pattern Recogn. 45, 6 (June 2012), 2237-2250. DOI=http://dx.doi.org/10.1016/j.patcog.2011.12.015
\bibitem{Ding:tri}
Chris Ding, Tao Li, Wei Peng, and Haesun Park. 2006. Orthogonal nonnegative matrix t-factorizations for clustering. In Proceedings of the 12th ACM SIGKDD international conference on Knowledge discovery and data mining (KDD '06). ACM, New York, NY, USA, 126-135. DOI=http://dx.doi.org/10.1145/1150402.1150420

\bibitem{cai:gnmf}
D. Cai, X. He, J. Han and T. S. Huang, Graph Regularized Nonnegative Matrix Factorization for Data Representation, in IEEE Transactions on Pattern Analysis and Machine Intelligence, vol. 33, no. 8, pp. 1548-1560, Aug. 2011. doi: 10.1109/TPAMI.2010.231
\bibitem{xu:document}
Wei Xu, Xin Liu, and Yihong Gong. 2003. Document clustering based on non-negative matrix factorization. In Proceedings of the 26th annual international ACM SIGIR conference on Research and development in informaion retrieval (SIGIR '03). ACM, New York, NY, USA, 267-273. DOI=http://dx.doi.org/10.1145/860435.860485
\bibitem{gu:localreg}
Zhou, Quanquan Gu Jie. Local learning regularized nonnegative matrix factorization. IJCAI 2009, Proceedings of the 21st International Joint Conference on Artificial Intelligence, Pasadena, California, USA, July 11-17, 2009
\bibitem{zhu:tri}
Linhong Zhu, Aram Galstyan, James Cheng, and Kristina Lerman. Tripartite graph clustering for dynamic sentiment analysis on social media. In Proceedings of the 2014 ACM SIGMOD International Conference on Management of Data, pages 1531-1542. ACM, 2014.

\end{thebibliography}
%

\appendix[Derivation of Equations \ref{eq:U1}, \ref{eq:H1}, \ref{eq:D1}, \ref{eq:W1}]\label{derivation}

To follow the conventional theory of constrained optimization we rewrite objective function \ref{eq:obj1} as;
\begin{align*}
\mathbf{J_{U,H,D,W}} &= Tr((\mathbf{X}_{uw} - \mathbf{UW}^T)(\mathbf{X}^{uw} - \mathbf{UW}^T)^T)\\ & + Tr((\mathbf{X}_{uh} - \mathbf{UH}^T)(\mathbf{X}_{uh} - \mathbf{UH}^T)^T)\\ & + Tr((\mathbf{X}_{ud} - \mathbf{UD}^T)(\mathbf{X}_{ud} - \mathbf{UD}^T)^T)\\ & + \alpha Tr(\mathbf{U}^TL_\mathbf{C}\mathbf{U}) + \gamma Tr(\mathbf{H}^TL_{\mathbf{H}_{sim}}\mathbf{H})\\ & + \theta Tr(\mathbf{D}^TL_{\mathbf{D}_{sim}}\mathbf{D}) + \beta Tr(\mathbf{W}^TL_{\mathbf{W}_{sim}}\mathbf{W}) \\
\mathbf{J_{U,H,D,W}} &= Tr(\mathbf{X}_{uw}\mathbf{X}^T_{uw}) - 2Tr(\mathbf{X}_{uw}\mathbf{WU}^T)\\ & + Tr(\mathbf{UW}^T\mathbf{WU}^T) + Tr(\mathbf{X}_{uh}\mathbf{X}^T_{uh})\\ & - 2Tr(\mathbf{X}_{uh}\mathbf{HU}^T) + Tr(\mathbf{UH}^T\mathbf{HU}^T)\\ & + Tr(\mathbf{X}_{ud}\mathbf{X}^T_{ud}) - 2Tr(\mathbf{X}_{ud}\mathbf{DU}^T) + Tr(\mathbf{UD}^T\mathbf{DU}^T) \\ &+ \alpha Tr(\mathbf{U}^TL_\mathbf{C}\mathbf{U}) + \gamma Tr(\mathbf{H}^TL_{\mathbf{H}_{sim}}\mathbf{H}) \\ &+ \theta Tr(\mathbf{D}^TL_{\mathbf{D}_{sim}}\mathbf{D}) + \beta Tr(\mathbf{W}^TL_{\mathbf{W}_{sim}}\mathbf{W}) 
\end{align*}
Let $\Phi$, $\eta$, $\Omega$ and $\Psi$ be the Lagrangian multipliers for constraints $\mathbf{U},\mathbf{H},\mathbf{D},\mathbf{W}>0$ respectively. So the Lagrangian function $ \mathcal{L}$ becomes;
\begin{align*}
\mathcal{L} &= Tr(\mathbf{X}_{uw}\mathbf{X}^T_{uw}) - 2Tr(\mathbf{X}_{uw}\mathbf{WU}^T) + Tr(\mathbf{UW}^T\mathbf{WU}^T)\\ & + Tr(\mathbf{X}_{uh}\mathbf{X}^T_{uh}) - 2Tr(\mathbf{X}_{uh}\mathbf{HU}^T) + Tr(\mathbf{UH}^T\mathbf{HU}^T)\\ & + Tr(\mathbf{X}_{ud}\mathbf{X}^T_{ud}) - 2Tr(\mathbf{X}_{ud}\mathbf{DU}^T) + Tr(\mathbf{UD}^T\mathbf{DU}^T)\\ & + \alpha Tr(\mathbf{U}^TL_\mathbf{C}\mathbf{U}) + \gamma Tr(\mathbf{H}^TL_{\mathbf{H}_{sim}}\mathbf{H}) + \theta Tr(\mathbf{D}^TL_{\mathbf{D}_{sim}}\mathbf{D})\\ & + \beta Tr(\mathbf{W}^TL_{\mathbf{W}_{sim}}\mathbf{W}) + Tr(\Phi \mathbf{U}^T) + Tr(\eta \mathbf{H}^T)\\ &  + Tr(\Omega \mathbf{D}^T) + Tr(\Psi \mathbf{W}^T) 
\end{align*}
The partial derivatives of Lagrangian function $\mathcal{L}$ with respect to $\mathbf{U}, \mathbf{H}, \mathbf{D}, \mathbf{W}$ are as follows;
\begin{align*}
\dfrac{\partial\mathcal{L}}{\partial \mathbf{U}} &= - 2\mathbf{X}_{uw}\mathbf{W} + 2\mathbf{UW}^T\mathbf{W} - 2\mathbf{X}_{uh}\mathbf{H} + 2\mathbf{UH}^T\mathbf{H} -\\ &\quad2\mathbf{X}_{ud}\mathbf{D} + 2\mathbf{UD}^T\mathbf{D} + 2\alpha L_\mathbf{C}\mathbf{U} + \Phi\\
\dfrac{\partial\mathcal{L}}{\partial \mathbf{H}} &= - 2\mathbf{X}_{uh}^T\mathbf{H} + 2\mathbf{UH}^T\mathbf{H} + 2\gamma L_{\mathbf{H}_{sim}}\mathbf{H} + \eta\\
\dfrac{\partial\mathcal{L}}{\partial \mathbf{D}} &= - 2\mathbf{X}_{ud}^T\mathbf{H} + 2\mathbf{UD}^T\mathbf{D} + 2\theta L_{\mathbf{D}_{sim}}\mathbf{D} + \Omega\\
\dfrac{\partial\mathcal{L}}{\partial \mathbf{W}} &= - 2\mathbf{X}_{uw}^T\mathbf{U} + 2\mathbf{WU}^T\mathbf{U} + 2\beta L_{\mathbf{W}_{sim}}\mathbf{W} + \Psi
\end{align*}
Setting derivatives equal to zero and using KKT complementarity conditions \cite{boyd:opt} of nonnegativity of matrices $\mathbf{U},\mathbf{H},\mathbf{D},\mathbf{W}$, $\Phi \mathbf{U} = 0$,  $\eta \mathbf{H} = 0$, $\Omega \mathbf{D} = 0$ and $\Psi \mathbf{W} = 0$, we get the update rules given in Equations \ref{eq:U1}, \ref{eq:H1}, \ref{eq:D1}, \ref{eq:W1}.
\end{document}